\documentclass{ifacconf}
\usepackage{cite}
\usepackage{amsmath,amssymb,amsfonts}
\usepackage{algorithmic}
\usepackage{graphicx}
\usepackage{textcomp}
\usepackage{xcolor}
\usepackage{mathrsfs}
\usepackage{psfrag}
\usepackage{cancel}
\usepackage{verbatim}
\usepackage{circuitikz}
\usetikzlibrary{decorations.pathreplacing,calligraphy,patterns}

\makeatletter
\let\old@ssect\@ssect 
\usepackage{natbib}
\usepackage{hyperref}
\hypersetup{colorlinks,linkcolor={blue},citecolor={blue},urlcolor={red}}
\def\@ssect#1#2#3#4#5#6{%
  \NR@gettitle{#6}
  \old@ssect{#1}{#2}{#3}{#4}{#5}{#6}
}
\makeatother

\def\mc{\mathcal}
\def\mb{\mathbb}

\newtheorem{definition}{Definition}
\newtheorem{theorem}{Theorem}

\newtheorem{assumption}{Assumption}
\newtheorem{proposition}{Proposition}

\begin{document}
\sloppy
\begin{frontmatter}
\title{\LARGE Negative Imaginary Neural ODEs: Learning to Control Mechanical Systems with Stability Guarantees\thanksref{footnoteinfo}}
\thanks[footnoteinfo]{This work was supported by the Australian Research Council (DP230101014).}
\author[ACFR]{Kanghong Shi}
\author[ACFR]{Ruigang Wang}
\author[ACFR]{Ian R. Manchester}

\address[ACFR]{Australian Centre for Robotics and School of Aerospace, Mechanical and Mechatronic Engineering, The University of Sydney, Sydney, NSW 2006, Australia. (Email: kanghong.shi@sydney.edu.au, ruigang.wang@sydney.edu.au, ian.manchester@sydney.edu.au)}

\maketitle
\thispagestyle{plain}
\pagestyle{plain}

\begin{abstract}
	We propose a neural control method to provide guaranteed stabilization for mechanical systems using a novel negative imaginary neural ordinary differential equation (NINODE) controller. Specifically, we employ neural networks with desired properties as state-space function matrices within a Hamiltonian framework to ensure the system possesses the NI property. This NINODE system can serve as a controller that asymptotically stabilizes an NI plant under certain conditions. For mechanical plants with colocated force actuators and position sensors, we demonstrate that all the conditions required for stability can be translated into regularity constraints on the neural networks used in the controller. We illustrate the utility, effectiveness, and stability guarantees of the NINODE controller through an example involving a nonlinear mass-spring system.
\end{abstract}
\begin{keyword}
negative imaginary systems, nonlinear systems, neural networks, neural ODEs, Hamiltonian systems.
\end{keyword}

\end{frontmatter}

\section{Introduction}
Modern control applications demand controllers that can handle highly nonlinear, high-dimensional, and uncertain dynamics, which might exceed the capability of classic model-based methods.  Because of the flexibility, expressiveness, and scalability of neural networks \cite{hornik1989multilayer}, they have been applied in system identification and control design since the 1990s; see e.g., \cite{miller1995neural,funahashi1993approximation,chen1992neural,ku1995diagonal}.
Recent research in neural networks has developed models that are more powerful and suitable for addressing control problems, see e.g. \cite{dawson2023safe}. 

However, the inherently black-box nature of neural networks complicates the derivation of formal stability and robustness guarantees, which are essential in safety-critical applications. Many techniques have been developed to embed certified control‑theoretic properties into neural network structures to address this challenge. For example, \cite{khansari2011learning} applies the Gaussian mixture model to construct and learn a stable system; \cite{kolter2019learning} learns a dynamics model together with its Lyapunov function to guarantee stability; \cite{revay2023recurrent} proposes the recurrent equilibrium neural networks to model contracting or incrementally passive discrete-time systems; \cite{greydanus2019hamiltonian} introduces Hamiltonian neural networks that obey the conservation of energy; see also \cite{dawson2023safe} for a survey on control with learned certificates.

Recently, the Polyak-\L{}ojasiewicz networks (PLNets) were introduced in \cite{wang2024monotone} and were applied in \cite{cheng2024learning} as the Hamiltonian to construct stable and passive neural ODEs. PLNets are constructed as the square of the Euclidean norm of a bi-Lipschitz network, and therefore possess both positive definiteness and the Polyak-\L{}ojasiewicz property \cite{wang2024monotone}. The level set of a PLNet is homeomorphic to a unit ball. A PLNet has a unique global minimum and hence is easily optimizable. Also, PLNets are non-convex, which makes them more flexible when serving as Lyapunov functions, compared to other convex neural networks such as the input convex neural networks used in \cite{kolter2019learning}. In this article, we apply PLNets to propose neural ODEs with certified negative imaginary (NI) property to control mechanical systems with colocated force actuators and position sensors.

Mechanical systems appear in a wide range of engineering applications such as robotics, vehicles, instruments, aerospace and manufacturing. However, controlling these systems can be challenging due to high-dimensional dynamics, nonlinearity, parameter uncertainties, underactuation, etc. These factors become particularly critical when it comes to flexible structures, which often exhibit high resonant dynamics resulting from insufficient internal damping \cite{preumont2018vibration}. In order to address the robust control problem for flexible structures, NI systems theory was introduced in \cite{lanzon2008stability,petersen2010feedback}.

The NI property characterizes the dissipative nature of mechanical systems equipped with colocated force actuators and position sensors; see also an interpretation for electrical circuits \cite{petersen2015physical}. Roughly speaking, a linear transfer function $G(s)$ is said to be NI if it is stable and its frequency response $G(j\omega)$ satisfies $j[G(j\omega)-G(j\omega)^*]\geq 0$ for all $\omega\geq 0$ \cite{lanzon2008stability}. Unlike passivity theory, which is limited to systems of relative degree zero or one, NI systems theory can deal with systems of relative degree zero, one or two \cite{shi2024necessary}. Additionally, while passivity theory uses negative velocity feedback control, NI systems theory uses positive position feedback control. Under certain conditions, the closed-loop interconnection of an NI system $G(s)$ and a strictly negative imaginary system $G_s(s)$ is asymptotically stable if and only if the DC loop gain is less than unity; i.e., $\lambda_{max}(G(0)G_s(0))<1$ \cite{lanzon2008stability}. 

Considering the nonlinear nature of most engineering systems, NI systems theory has been extended to nonlinear systems \cite{ghallab2018extending,shi2021robust,shi2023output} via the related notion of counterclockwise dynamics \cite{angeli2006systems}. A nonlinear system is said to be NI if it is dissipative with respect to the supply rate $u^T\dot y$, where $u$ and $y$ are the input and output of the system, respectively. Similar to linear NI systems theory, a nonlinear NI system can be stabilized using an NI controller with some strictness, given that some assumptions are satisfied. NI systems theory has been applied in many fields such as nano-positioning control \cite{mabrok2013spectral}, the control of lightly damped structures \cite{bhikkaji2011negative}, and the control of electrical power systems \cite{chen2024nonlinear}.

Despite its theoretical advancements and broad applicability, existing literature on NI systems theory lacks a systematic methodology for controller synthesis. Given an NI plant, there is no established procedure to identify an optimal controller model, and each candidate controller requires individual verification to ensure that the stability conditions are satisfied. This makes the controller design difficult, especially for plants with large input and output dimensions. In this work, we address these difficulties by introducing a novel negative imaginary ordinary differential equation (NINODE) controller, which has an intrinsic NI property. Such a NINODE controller can guarantee asymptotic stability when applied to a mechanical system with NI properties.

Specifically, we use a particular Hamiltonian form to represent NI systems, as introduced in \cite{van2011positive,van2016interconnections}, termed `input-output Hamiltonian systems with dissipation'. This model ensures the NI property provided that the function matrices within the model satisfy some symmetry or positive definiteness conditions. To enhance the flexibility of the system model, we parameterize these function matrices using neural networks with certified properties; see \cite{wang2024monotone} for details of such neural networks. The inherent structures of these neural networks guarantee their properties, thereby allowing direct parameterization of the controller model without imposing additional constraints. Similar methodologies have been applied in the construction of Lyapunov stable and passive systems; see \cite{cheng2024learning} for details and a discussion on the advantages. The current work can be regarded as a complementary result to these previous passivity-based neural control results, as it is applicable to plants with position sensors. In addition, we provide a closed-loop stability analysis for the interconnection of a mechanical plant and a NINODE controller.

In this paper, we first show that the interconnection of two NI systems in Hamiltonian form, with one having some strictness, is asymptotically stable under certain assumptions. Specifically, for a mechanical plant, we show that the required assumptions can be formulated as regularity constraints on the neural networks within the controller model. The Hamiltonian of the NI controller, which also serves as its storage function, is constructed using PLNets \cite{wang2024monotone}, which possess the Polyak-\L{}ojasiewicz (PL) properties and are therefore positive definite. Additionally, the output function is constructed as the first $p$ rows of bi-Lipschitz networks \cite{wang2024monotone}, where $p$ denotes the system's output dimension. The regularity parameters of these neural networks only need to satisfy a simple inequality involving the stiffness matrix of the mechanical plant.

The rest of the paper is organized as follows: Section \ref{sec:Pre} provides essential definitions of NI systems, PLNets and bi-Lipschitz nets, which are later used in the paper. Section \ref{sec:NI Hamiltonian} presents the stability result for the interconnection of two NI systems in the Hamiltonian form. Section \ref{sec:Mechanical} provides the main results of the paper. For a mechanical system, we propose an NI controller constructed using neural networks, and we prove the closed-loop stability. Section \ref{sec:example} illustrates the utility and advantages of our proposed control methodology on a nonlinear mass-spring system.

Notation: The notation in this paper is standard. $\mathbb R$ denotes the set of real numbers. $\mb N_+$ denotes the set of positive integers. $\mathbb R^{m\times n}$ denotes the space of real matrices of dimension $m\times n$. $A^T$ and $A^*$ denote the transpose and the complex conjugate transpose of a matrix $A$, respectively. $A^{-T}$ denotes the transpose of the inverse of $A$; that is, $A^{-T}=(A^{-1})^T=(A^T)^{-1}$. $\lambda_{max}(A)$ denotes the largest eigenvalue of a matrix $A$ with real spectrum. $|A|_{op}$ denotes the operator norm of a matrix. $|\cdot|$ denotes the standard Euclidean norm. For a real symmetric or complex Hermitian matrix $P$, $P>0\ (P\geq 0)$ denotes the positive (semi-)definiteness of a matrix $P$ and $P<0\ (P\leq 0)$ denotes the negative (semi-)definiteness of a matrix $P$. A function $V: \mb R^n \to \mb R$ is called positive definite if $V(0)=0$ and $V(x)>0$ for all $x\neq 0$. A function $V: \mb R^n \to \mb R$ is said to be of class $C^1$ if it is continuously differentiable; $V$ is said to be of class $C^2$ if it is twice continuously differentiable. $\nabla$ denotes the gradient operator.

\section{Preliminaries}\label{sec:Pre}
\subsection{Negative imaginary systems}
Consider a nonlinear system with the following model:
\begin{subequations}\label{eq:general nonlinear system}
	\begin{align}
    \dot x(t)=&\ f(x(t),u(t)),\label{eq:state equation of nonlinear NI}\\
    y(t)=&\ h(x(t)),
    \label{eq:output equation of nonlinear NI}
\end{align}
\end{subequations}
where $x(t)\in \mathbb R^{n}$ is the state, $u(t)\in \mathbb R^p$ is a locally integrable input, and $y(t)\in \mathbb R^p$ is the output, $f:\mathbb R^n\times \mathbb R^p \to \mathbb R^n$ is a Lipschitz continuous function and $h:\mathbb R^n \to \mathbb R^p$ is a class $C^1$ function.
\begin{definition}\label{def:nonlinear NI}\cite{ghallab2018extending,shi2021robust}
The system (\ref{eq:general nonlinear system}) is said to be a nonlinear negative imaginary (NI) system if there exists a positive definite storage function $V:\mathbb R^n\to \mathbb R$ of class $C^1$ such that for any locally integrable input $u$ and any solution $x$ to (\ref{eq:general nonlinear system}),
\begin{equation}\label{eq:NI MIMO definition inequality}
    \dot V(x(t))\leq u(t)^T\dot y(t), \quad \forall t\geq 0.
\end{equation}
\end{definition}
\subsection{Bi-Lipschitz and Polyak-\L{}ojasiewicz Neural Networks}
\begin{definition}
	A mapping $\mc G(x):\mb R^n\to \mb R^n$ is said to be bi-Lipschitz if there exists $\nu\geq \mu>0$ such that for all $x_a,x_b\in \mb R^n$,
	\begin{equation}
		\mu|x_a-x_b|\leq |\mc G(x_a)-\mc  G (x_b)|\leq \nu |x_a-x_b|.
	\end{equation}
We also say that $\mc G(x)$ is $(\mu,\nu)$-Lipschitz.
\end{definition}

\begin{definition}
	A neural network $H:\mb R^n \to \mb R$ is said to be a Polyak-\L{}ojasiewicz Network (PLNet) if for all $x\in \mb R^n$,
	\begin{equation}\label{eq:PLNet}
		\frac{1}{2}|\nabla_x H(x)|^2 \geq \varrho \left|H(x)-\min_x H(x)\right|,
	\end{equation}
where $\varrho>0$.
\end{definition}

\begin{proposition}\cite{wang2024monotone}
Suppose $\mc G(x):\mb R^n \to \mb R^n$	is $\mu$-inverse Lipschitz; i.e., $\mu|x_a-x_b|\leq |\mc G(x_a)-\mc G(x_b)|$ for all $x_a,x_b\in \mb R^n$. Then
\begin{equation}
	H(x) = \frac{1}{2}|\mc G(x)|^2 + c,
\end{equation}
with $c>0$ a scalar, is a PLNet that satisfies (\ref{eq:PLNet}) with $\varrho = \mu^2$.
\end{proposition}

\section{Control of an NI Hamiltonian system}\label{sec:NI Hamiltonian}
While port-Hamiltonian systems have the passivity property, \cite{van2016interconnections,van2011positive} also introduce a class of Hamiltonian systems that have the NI property. Consider a Hamiltonian system of the following form
\begin{subequations}\label{eq:Hamiltonian}
	\begin{align}
	\mc H_1:\quad	\dot x_1 =&\ [J_1(x_1)-R_1(x_1)]\left(\nabla H_1(x_1)-\nabla^T C_1(x_1)u_1\right),\label{eq:pH1 state}\\
		y_1 =&\ C_1(x_1),\label{eq:pH1 output}
	\end{align}
\end{subequations}
where $x_1\in \mb R^n$, $u_1,y_1\in \mb R^p$ are the state, input and output of the system, respectively. Here, $J_1:\mb R^n\to \mb R^n$ is skew-symmetric; i.e., $J_1(x_1) = -J_1(x_1)^T$. Also, $R_1:\mb R^n\to \mb R^n$ is symmetric; i.e., $R_1(x_1) = R_1(x_1)^T$. Here,  $C_1:\mb R^n \to \mb R^p$ is a class $C^1$ function and $C_1(0)=0$.
\begin{theorem}(see \cite{van2016interconnections,van2011positive})
	Suppose $H_1(x_1)$ is a positive definite and $R_1(x_1)$ is a positive semidefinite matrix for all $x_1\in \mb R^n$, then the system (\ref{eq:Hamiltonian}) is negative imaginary with the storage function $H_1(x_1)$.
\end{theorem}

Also, consider a Hamiltonian system of a similar form:
\begin{subequations}\label{eq:Hamiltonian2}
	\begin{align}
	\mc H_2:\	\dot x_2 =&\ [J_2(x_2)-R_2(x_2)]\left(\nabla H_2(x_2)-\nabla^T C_2(x_2)u_2\right),\label{eq:pH2 state}\\
		y_2 =&\ C_2(x_2),\label{eq:pH2 output}
	\end{align}
\end{subequations}
where $x_2\in \mb R^m$, $u_2,y_2\in \mb R^p$ are the state, input and output of the system, respectively. Here, $J_2:\mb R^m\to \mb R^m$ is skew-symmetric; i.e., $J_2(x_2) = -J_2(x_2)^T$. Also, $R_2:\mb R^m\to \mb R^m$ is symmetric; i.e., $R_2(x_2) = R_2(x_2)^T$. Here,  $C_2:\mb R^m \to \mb R^p$ is a class $C^1$ function and $C_2(0)=0$.

\begin{figure}[h!]
\centering
\psfrag{in_1}{$u_1$}
\psfrag{y_1}{$y_1$}
\psfrag{in_2}{$u_2$}
\psfrag{y_2}{$y_2$}
\psfrag{plant}{\hspace{0cm}$\mc H_1$}
\psfrag{con}{\hspace{0.1cm}$\mc H_2$}
\includegraphics[width=8.5cm]{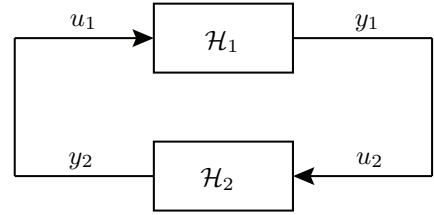}
\caption{Closed-loop interconnection of the systems $\mc H_1$ and $\mc H_2$ given in (\ref{eq:Hamiltonian}) and (\ref{eq:Hamiltonian2}), respectively.}
\label{fig:interconnection}
\end{figure}

We investigate in the following the stability for the interconnection of the systems $\mc H_1$ and $\mc H_2$ as shown in Fig.~\ref{fig:interconnection} with
\begin{equation}
    u_1 =\ y_2,\quad u_2 =\ y_1.\label{eq:setting}
\end{equation}

We require the following assumptions to be satisfied.

\begin{assumption}\label{assumption1}
	The system $\mc H_1$ given in (\ref{eq:Hamiltonian}) is such that over any time interval $[t_a,t_b]$ where $t_a<t_b$, $\dot C_1(x_1)\equiv 0$ if and only if $\dot x_1\equiv 0$.
\end{assumption}

\begin{assumption}\label{assumption2}
	The system $\mc H_2$ given in (\ref{eq:Hamiltonian2}) is such that over any time interval $[t_a,t_b]$ where $t_a<t_b$, if $\nabla H_2(x_2)-\nabla^T C_2(x_2)u_2\equiv 0$, then $u_2$ remains constant over this time interval.
\end{assumption}

\begin{assumption}\label{assumption3}
	The equations
	\begin{align}
		[J_1(x_1)-R_1(x_1)]\left(\nabla H_1(x_1)-\nabla^T C_1(x_1)C_2(x_2)\right)=&\ 0;\label{eq:assumption3 eq1}\\
		\nabla H_2(x_2)-\nabla^T C_2(x_2)C_1(x_1)=&\ 0\notag
	\end{align}
has only one common solution at $(x_1,x_2)=(0,0)$.
\end{assumption}

Assumption \ref{assumption1} is an observability assumption. Assumption \ref{assumption2} requires all the inputs to affect the system dynamics. Assumption \ref{assumption3} rules out the case that the cascade of $\mc H_1$ and $\mc H_2$ has the DC gain $1$. Assumption \ref{assumption3}, together with the requirement of a positive definite candidate Lyapunov function in Theorem \ref{theorem1}, can be regarded as the nonlinear counterpart of the DC gain condition required in \cite{lanzon2008stability} for linear NI systems stability. Note that Assumptions \ref{assumption2} and \ref{assumption3} only pose restrictions on the controller $\mc H_2$.

\begin{theorem}\label{theorem1}
	Consider the positive feedback interconnection of the systems (\ref{eq:Hamiltonian}) and (\ref{eq:Hamiltonian2}) as shown in Fig.~\ref{fig:interconnection}. Suppose $R_1(x_1)\geq 0$ for all $x_1\in \mb R^n$, $R_2(x_2)>0$ for all $x_2\in \mb R^m$, and Assumptions \ref{assumption1}, \ref{assumption2} and \ref{assumption3} are satisfied. Also, suppose the function
	\begin{equation}\label{eq:W}
		W(x_1,x_2) = H_1(x_1)+H_2(x_2)-C_1(x_1)^TC_2(x_2)
	\end{equation}
is positive definite. Then the closed-loop system shown in Fig.~\ref{fig:interconnection} is Lyapunov stable.
\end{theorem}
\begin{pf}
	We prove stability for the closed-loop system using Lyapunov's direct method, with $W(x_1,x_2)$ given in (\ref{eq:W}) used as the candidate Lyapunov function. Taking the time derivative of $W$, we have
	\begin{align}
		\dot W&(x_1,x_2)\notag\\
		 =&\ \nabla^T H_1(x_1)\dot x_1+\nabla^T H_2(x_2)\dot x_2-C_1(x_1)^T\nabla C_2(x_2)\dot x_2\notag\\
		 &-C_2(x_2)^T\nabla C_1(x_1)\dot x_1\notag\\
		=& \left[\nabla^T H_1(x_1)-C_2(x_2)^T\nabla C_1(x_1)\right]\dot x_1\notag\\
		&+\left[\nabla^T H_2(x_2)-C_1(x_1)^T\nabla C_2(x_2)\right]\dot x_2\notag\\
		=& \left[\nabla^T H_1(x_1)-C_2(x_2)^T\nabla C_1(x_1)\right][J_1(x_1)-R_1(x_1)]\cdot\notag\\&\left[\nabla H_1(x_1)-\nabla^T C_1(x_1)C_2(x_2)\right]\notag\\
		&+\left[\nabla^T H_2(x_2)-C_1(x_1)^T\nabla C_2(x_2)\right][J_2(x_2)-R_2(x_2)]\cdot\notag\\
		&\left[\nabla H_2(x_2)-\nabla^T C_2(x_2)C_1(x_1)\right]\notag\\
		=& -\left[\nabla^T H_1(x_1)-C_2(x_2)^T\nabla C_1(x_1)\right]R_1(x_1)\cdot \notag\\
		&\left[\nabla H_1(x_1)-\nabla^T C_1(x_1)C_2(x_2)\right]\notag\\
		&-\left[\nabla^T H_2(x_2)-C_1(x_1)^T\nabla C_2(x_2)\right]R_2(x_2)\cdot \notag\\
		&\left[\nabla H_2(x_2)-\nabla^T C_2(x_2)C_1(x_1)\right]
		\leq 0,\notag
	\end{align}
	where (\ref{eq:pH1 output}), (\ref{eq:pH2 output}) and (\ref{eq:setting}) are used. The system is stable in the sense of Lyapunov. We apply LaSalle's invariance principle to prove asymptotic stability. When $\dot W(x_1,x_2)\equiv 0$, we have $\nabla H_2(x_2)-\nabla^T C_2(x_2)C_1(x_1) \equiv 0$ because $R_2(x_2)>0$ for all $x_2\in \mb R^m$. Then, $\dot x_2 \equiv 0$ according to (\ref{eq:pH2 state}). According to Assumption \ref{assumption2}, we have that $u_2=C_1(x_1)$ remains constant. Hence, $x_1$ remains constant, according to Assumption \ref{assumption1}. This implies that $\dot x_1\equiv 0$, and hence (\ref{eq:assumption3 eq1}) holds over time. According to Assumption \ref{assumption3}, this is only possible if $x_1=0$ and $x_2=0$. Otherwise, $\dot W(x_1,x_2)$ cannot stay at zero, and $W(x_1,x_2)$ will keep decreasing until the system reaches its equilibrium. \hfill $\blacksquare$
\end{pf}

\section{The Mechanical System Case}\label{sec:Mechanical}
In this section, we specialize the results in Section \ref{sec:NI Hamiltonian} to fully actuated mechanical systems. We consider a special form of system (\ref{eq:Hamiltonian}) given as follows, which describes the motion of a mechanical system with position output.
\begin{subequations}\label{eq:mechanical pH}
	\begin{align}
\dot x_1 =& \left(\begin{bmatrix}
			0 & I\\-I & 0
		\end{bmatrix}-\begin{bmatrix}
			0 & 0\\0 & r_1(x_1)
		\end{bmatrix}\right)\cdot\notag\\
		&\left(\begin{bmatrix}
			\nabla V(q)+\frac{1}{2}\nabla_q\left(p^TM(q)^{-1}p\right)\\M(q)^{-1}p
		\end{bmatrix}- \begin{bmatrix}
			I\\0
		\end{bmatrix} u_1\right)\label{eq:mechanical pH state},\\
		y_1 =& \begin{bmatrix}
			I & 0
		\end{bmatrix}x_1.
	\end{align}
\end{subequations} 
Here, the system state is $x_1=\begin{bmatrix}
	q^T & p^T
\end{bmatrix}^T$, where $q\in \mathbb R^n$ denotes the position coordinates, and $p=M(q)\dot q \in \mathbb R^n$ denotes the generalized momenta. Also, $M(q)=M(q)^T\in \mathbb R^{n\times n}$ is the mass function matrix, which is positive definite for all $q\in \mb R^n$. The function matrix $r_1(x_1)\in\mathbb R^{n\times n}$ describes the damping effects in the system and is assumed to be symmetric and positive semidefinite for all $x_1\in \mb R^{2n}$. The Hamiltonian of the system is
\begin{equation}\label{eq:Hamiltonian of mechanical system}
	H_1(x_1) = V(q)+\frac{1}{2}p^TM(q)^{-1}p,
\end{equation}
where $V(q)$ is a class $C^1$ function that denotes the system's potential energy. We assume that $V(q)$ is positive definite and satisfies the Polyak-\L{}ojasiewicz condition; i.e.,
\begin{align}
    V(0)=0, \quad V(q) \geq \frac{1}{2}\eta |q|^2 &\text{ for all } q\in\mb R^n\backslash\{0\};\label{eq:V(q) PD}\\
    \frac{1}{2}|\nabla V(q)|^2 \geq \eta V(q) &\text{ for all } q \in \mathcal \mb R^n,\label{eq:V(q) PL}
\end{align}
where $\eta>0$ is a scalar. The system model (\ref{eq:mechanical pH}) is obtained by substituting $J_1(x_1)=\begin{bmatrix}
			0 & I\\-I & 0
		\end{bmatrix}$, $R_1(x_1)=\begin{bmatrix}
			0 & 0\\0 & r_1(x_1)
		\end{bmatrix}$, $C_1(x_1)=\begin{bmatrix}
			I & 0
		\end{bmatrix}x_1$ and (\ref{eq:Hamiltonian of mechanical system}) in (\ref{eq:Hamiltonian}).

We apply a controller of a form similar to (\ref{eq:Hamiltonian2}). However, besides the measured output $y_1$ of the plant, we allow the controller dynamics to be also influenced by some other sensor data, which we denote by $z(t)\in \mb R^l$; e.g., data from a video camera. The model of the controller is given as follows.
\begin{subequations}\label{eq:controller}
	\begin{align}
	\dot x_2 =&\ [J_2(x_2,z)-R_2(x_2,z)]\left(\nabla H_2(x_2)-\nabla^T C_2(x_2)u_2\right),\label{eq:controller state}\\
		y_2 =&\ C_2(x_2),\label{eq:controller output}
	\end{align}
\end{subequations}
where $x_2\in \mb R^m$, $u_2,y_2\in \mb R^n$ $(m\geq n)$ are the state, input and output of the system, respectively. Here, $J_2:\mb R^m\times \mb R^l\to \mb R^m$ is skew-symmetric; i.e., $J_2(x_2,z) = -J_2(x_2,z)^T$ for all $(x_2,z)$. Also, $R_2:\mb R^m\times \mb R^l\to \mb R^m$ is symmetric and positive definite; i.e., $R_2(x_2,z) = R_2(x_2,z)^T>0$ for all $(x_2,z)$. Here,  $C_2:\mb R^m \to \mb R^n$ is a class $C^1$ function.

\begin{theorem}\label{theorem:mechanical system}
    Consider the closed-loop interconnection of the plant (\ref{eq:mechanical pH}) and the controller (\ref{eq:controller}) as shown in Fig.~\ref{fig:interconnection}. Suppose the controller Hamiltonian is a PLNet given by
\begin{equation}\label{eq:PLNet H_2}
		H_2(x_2) = \frac{1}{2}|\mc G(x_2)|^2,
\end{equation}
where $\mc G(x_2)$ is $\gamma$-inverse Lipschitz and $\mc G(0)=0$. Also, suppose the controller's output equation is given by
	\begin{equation}\label{eq:C_2 parameterization}
		C_2(x_2)=\begin{bmatrix}
		I_n & {\bf 0}_{n\times (m-n)}
	\end{bmatrix}\overline C_2(x_2),
	\end{equation}
where $\overline C_2(x_2):\mb R^m\to \mb R^m$ is a $(\underline \mu,\overline \mu)$-Lipschitz Net, with $0<\underline \mu<\overline \mu<\sqrt{\eta}\gamma$ and  $\overline C_2(0)=0$. Then the closed-loop system is asymptotically stable.
\end{theorem}

\begin{pf}
	We apply Lyapunov's direct method using the following candidate Lyapunov function:
	\begin{align}
		W(x_1,x_2)=&\ H_1(x_1)+H_2(x_2)-C_1(x_1)^TC_2(x_2)\notag\\
		 =&\ V(q)+\frac{1}{2}p^TM(q)^{-1}p+H_2(x_2)-q^TC_2(x_2)\notag\\
		\geq &\ \frac{1}{2}\eta |q|^2+\frac{1}{2}p^TM(q)^{-1}p+H_2(x_2)\notag\\
		&-q^TC_2(x_2).\notag
	\end{align}
First, we show that the function $W(x_1,x_2)$ is positive definite. It can be observed that $W(0,0)=0$. Now, we show that $W(x_1,x_2)>0$ for all $(x_1,x_2)\neq 0$. According to (\ref{eq:PLNet H_2}), we have that
\begin{equation*}
	H_2(x_2)\geq \frac{1}{2}\gamma^2 |x_2|^2.
\end{equation*}
Also, since $\overline C_2(x_2)$ is $(\underline \mu,\overline \mu)$-Lipschitz, then
\begin{equation*}
	\left|\overline C_2(x_2)\right|\leq \overline \mu |x_2|.
\end{equation*}
Therefore, according to (\ref{eq:C_2 parameterization}), we have
\begin{equation*}
	|C_2(x_2)|\leq \left|\overline C_2(x_2)\right|\leq \overline \mu |x_2|.
\end{equation*}
Then, we have that
\begin{align}
	q^TC_2(x_2) =&\ |q||C_2(x_2)|\cos\langle q,C_2(x_2) \rangle\leq |q||C_2(x_2)|\notag\\
	&\leq \overline \mu |q||x_2|.\notag
\end{align}
Therefore,
\begin{align}
	\frac{1}{2}\eta |q|^2 &+H_2(x_2)-q^TC_2(x_2)\notag\\
	&\geq \frac{1}{2}\eta |q|^2+\frac{1}{2}\gamma^2|x_2|^2-\overline \mu |q||x_2|,\notag
\end{align}
which is positive definite with respect to $(q,x_2)$ because $\overline \mu^2-\eta \gamma^2<0$. Also, since $p^TM(q)^{-1}p>0$ for all $p\neq 0$, then the function $W(x_1,x_2)$ is positive definite. Also, $W(x_1,x_2)$ is radially unbounded. Taking the time derivative of $W(x_1,x_2)$, we have that
\begin{align}
	\dot W(& x_1,x_2)\notag\\
	=&\ \nabla^T V(q)\dot q+p^TM(q)^{-1}\dot p + \frac{1}{2}\nabla_q\left(p^TM(q)^{-1}p\right)\dot q \notag\\
	& + \dot H_2(x_2)-p^TM(q)^{-1}C_2(x_2)-q^T\dot C_2(x_2)\notag\\
	=&\ \nabla^T V(q) M(q)^{-1}p+p^TM(q)^{-1}\cdot\notag\\
	&\left(-\nabla V(q)-\frac{1}{2}\nabla_q\left(p^TM(q)^{-1}p\right)-r_1(x_1)M(q)^{-1}p\right.\notag\\
	&+C_2(x_2)\bigg)+ \frac{1}{2}\nabla_q\left(p^TM(q)^{-1}p\right)M(q)^{-1}p+\dot H_2(x_2)\notag\\
	&-p^TM(q)^{-1}C_2(x_2)-q^T\dot C_2(x_2)\notag\\
    =&\ -p^TM(q)^{-1}r_1(x_1)M(q)^{-1}p+\dot H_2(x_2)-q^T\dot C_2(x_2)\notag\\
	=&\ -p^TM(q)^{-1}r_1(x_1)M(q)^{-1}p\notag\\
	&-\left[\nabla^T H_2(x_2)-q^T\nabla C_2(x_2)\right]R_2(x_2,z)\cdot\notag\\
	&\left[\nabla H_2(x_2)-\nabla^T C_2(x_2)q\right] \leq 0.\notag
\end{align}
The system is Lyapunov stable. Now we apply LaSalle's invariance principle to show asymptotic stability. In the case that $\dot W(x_1,x_2)\equiv 0$, we have
\begin{equation}\label{eq:controller steady state}
	\nabla H_2(x_2)-\nabla^T C_2(x_2)q\equiv 0.
\end{equation}
This implies that $\dot x_2\equiv 0$ according to (\ref{eq:controller state}). In this case, the controller (\ref{eq:controller}) reaches a steady state. Also, since $\overline C_2(x_2)$ is bi-Lipschitz, then $\nabla \overline C_2(x_2)$ is invertible for all $x_2\in \mb R^m$. Considering (\ref{eq:C_2 parameterization}), we have
\begin{equation*}
	\nabla^TC_2(x_2)=\nabla^T \overline C_2(x_2)\begin{bmatrix}
		I_n \\ {\bf 0}_{(m-n)\times n}
	\end{bmatrix}.
\end{equation*}
This implies that $\nabla^TC_2(x_2)$ has full column rank. This, together with the fact that $\dot x_2 \equiv 0$, implies that $\dot q\equiv 0$ according to (\ref{eq:controller steady state}). Therefore, we have that $p\equiv 0$ according to (\ref{eq:mechanical pH state}). This implies that the plant (\ref{eq:mechanical pH}) also reaches a steady state. Then, according to (\ref{eq:mechanical pH state}), we have
\begin{equation}\label{eq:plant steady state}
	u_1 \equiv y_2 \equiv C_2(x_2) \equiv \nabla V(q).
\end{equation}
Note that $\nabla_q(p^TM(q)^{-1}p)=0$ in this case because $p\equiv 0$. We show in the following that the two conditions (\ref{eq:controller steady state}) and (\ref{eq:plant steady state}), which describe the relationship between $x_2$ and $q$ cannot hold together except for the case that $x_2=0$ and $q=0$. Since $H_2(x_2)$ is a PLNet and $H_2(0)=0$, then we have that
\begin{equation*}
    \frac{1}{2}|\nabla H_2(x_2)|^2\geq \gamma ^2 |H_2(x_2)| = \frac{1}{2}\gamma^2 |\mathcal G(x_2)|^2.
\end{equation*}
Therefore,
\begin{equation}\label{eq:nabla H_2 bound}
    |\nabla H_2(x_2)|\geq \gamma |\mathcal G(x_2)|\geq \gamma^2 |x_2|,
\end{equation}
where the last inequality uses the inverse Lipschitz property of $\mathcal G(x_2)$. Considering the $(\underline \mu,\overline \mu)$-Lipschitz property of $\overline C_2(x_2)$, (\ref{eq:C_2 parameterization}), and the fact that $\overline C_2(0)=0$ we have
\begin{equation}\label{eq:C_2 bound}
    |C_2(x_2)| \leq |\overline C_2(x_2)|\leq \overline \mu |x_2|.
\end{equation}
Therefore, $C_2(x_2)$ is also $\overline \mu$-Lipschitz. The operator norm of $\nabla C_2(x_2)$ is bounded by the Lipschitz constant of $C_2(x_2)$; i.e., $|\nabla C_2(x_2)|_{op}\leq \overline \mu$. Therefore,
\begin{equation*}
   |\nabla^T C_2(x_2)|_{op} = |\nabla C_2(x_2)|_{op}\leq \overline \mu.
\end{equation*}
Hence, we have that
\begin{equation}\label{eq:nabla C_2 q bound}
    |\nabla^T C_2(x_2) q|\leq |\nabla^T C_2(x_2)|_{op} |q| \leq \overline \mu |q|.
\end{equation}
Considering (\ref{eq:controller steady state}), (\ref{eq:nabla H_2 bound}) and (\ref{eq:nabla C_2 q bound}), we have that
\begin{equation}\label{eq:x_2 leq xx q}
    \gamma^2|x_2|\leq |\nabla H_2(x_2)|=|\nabla^T C_2(x_2)q|\leq \overline \mu |q|.
\end{equation}
Also, according to (\ref{eq:V(q) PD}) and (\ref{eq:V(q) PL}), we have that
\begin{equation*}
    |\nabla V(q)|^2 \geq 2 \eta V(q) \geq \eta^2|q|^2,
\end{equation*}
which implies
\begin{equation} \label{eq:|q| upper bound}
    |\nabla V(q)| \geq \eta |q|.
\end{equation}
Also, considering (\ref{eq:plant steady state}), (\ref{eq:C_2 bound}) and (\ref{eq:|q| upper bound}), we have that
\begin{equation} \label{eq:x_2 geq xx q}
    \overline \mu |x_2|\geq |C_2(x_2)|=|\nabla V(q)|\geq \eta |q|.
\end{equation}
Combining (\ref{eq:x_2 leq xx q}) and (\ref{eq:x_2 geq xx q}), we have that
\begin{equation*}
    \frac{\eta}{\overline \mu}|q| \leq |x_2| \leq \frac{\overline \mu}{\gamma^2}|q|,
\end{equation*}
which have nonzero solutions of $|x_2|$ and $|q|$ only if $\overline \mu \geq \sqrt{\eta}\gamma$. This contradicts the parameter constraints $\overline \mu < \sqrt{\eta}\gamma$. Therefore, $\dot W(x_1,x_2)$ cannot stay at zero and will keep decreasing until the states reach the equilibrium. This implies that the system is asymptotically stable. \hfill $\blacksquare$
\end{pf}

\section{Illustrative Example}\label{sec:example}
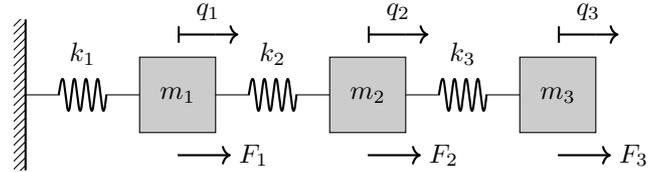
\begin{figure}[h!]
\centering
\ctikzset{bipoles/length=1.2cm}
\begin{circuitikz}[BC/.style = {decorate,
    decoration={calligraphic brace, mirror, amplitude=2.5mm,
    raise=1mm},
    very thick, pen colour=black}]
   \pattern[pattern=north east lines] (-0.15,0) rectangle (0,2);
    \draw[thick] (0,0) -- (0,2);
    \draw (0,1) to [spring, l = $k_1$] (1.5,1);
    \draw[fill=gray!40] (1.5,0.5) rectangle (2.5,1.5);
    \node at (2,1){$m_1$};
    \draw (2.5,1) to [spring, l = $k_2$] (4,1);
    \draw[fill=gray!40] (4,0.5) rectangle (5,1.5);
    \node at (4.5,1){$m_2$};
    \draw (5,1) to [spring, l = $k_3$] (6.5,1);
    \draw[fill=gray!40] (6.5,0.5) rectangle (7.5,1.5);
    \node at (7,1){$m_3$};
    \draw[thick, ->] (2,0.2) -- (2.7,0.2);
    \node at (3,0.2){$F_1$};    
    \draw[thick, ->] (4.5,0.2) -- (5.2,0.2);
    \node at (5.5,0.2){$F_2$};   
    \draw[thick, ->] (7,0.2) -- (7.7,0.2);
    \node at (8,0.2){$F_3$};
    \draw[thick, |-] (2,1.8) -- (2.5,1.8);
    \draw[thick, ->] (2.5,1.8) -- (2.8,1.8);
    \node at (2.4,2.1){$q_1$};   
    \draw[thick, |-] (4.5,1.8) -- (5,1.8);
    \draw[thick, ->] (5,1.8) -- (5.3,1.8);
    \node at (4.9,2.1){$q_2$};    
    \draw[thick, |-] (7,1.8) -- (7.5,1.8);
    \draw[thick, ->] (7.5,1.8) -- (7.8,1.8);
    \node at (7.4,2.1){$q_3$};
\end{circuitikz}
\caption{The top view of a mass-spring system containing three masses, which move rectilinearly on a frictionless floor. All the springs are nonlinear. The displacements of the masses are denoted by $q_1$, $q_2$ and $q_3$, respectively. We apply external forces to the masses and denote them by $F_1$, $F_2$ and $F_3$, respectively.}
\label{fig:example}
\end{figure}

We provide an example to illustrate the utility of the proposed control approach and its advantages. To be specific, we consider stabilizing the mass-spring system shown in Fig.~\ref{fig:example} while also minimizing a certain cost function. The masses are cascaded in series with nonlinear springs. External control forces are applied to the masses.

The parameters in the mass-spring system are given as follows. The masses are
$$m_1 = 0.02~\mathrm{kg}, \quad m_2 = 0.01~\mathrm{kg}, \quad m_3 = 0.03~\mathrm{kg}.$$
The nonlinear spring force $F_{si}$ from the spring $k_i$ ($i=1,2,3$) is given by
$$F_{si} = k_{il}\Delta_{i} + k_{in}\Delta_{i}^3,$$
where $k_{il}$ and $k_{in}$ denote the linear and nonlinear spring constants of the spring $k_i$, respectively. Here, $\Delta_i$ denotes the displacement that the spring $k_i$ is compressed from its equilibrium length. The spring constants are
$$k_{1l} = 15, \quad k_{2l} = 10, \quad k_{3l} = 20;$$
$$k_{1n} = 5, \quad k_{2n} = 2, \quad k_{3n} = 3.$$
Denote the displacements of the masses $m_1$, $m_2$ and $m_3$ from the positions where the springs are unstretched by $q_1$, $q_2$ and $q_3$. Denote the generalized momenta of the masses by $p_1$, $p_2$ and $p_3$, respectively. That is, $p_i = m_i \dot q_i$ for $i=1,2,3$.

We aim to construct a learnable controller of the form (\ref{eq:controller}) to generate the control input
\begin{equation*}
	u_1 = \begin{bmatrix}
			F_1&F_2&F_3
		\end{bmatrix}^T
\end{equation*}
for the mass-spring plant. We preset constraints to the controller as specified in Theorem \ref{theorem:mechanical system} so that closed-loop asymptotic stability is guaranteed. In addition, we train the functions $J_2$, $R_2$, $C_2$ and $H_2$ in the controller (\ref{eq:controller}) to minimize the following quadratic cost function:
\begin{equation}\label{eq:loss function J}
	\mc J = \sum_{k=1}^{N_s} \left(x_k^T\mc Qx_k+u_k^T\mc R u_k\right),
\end{equation}
which is also referred to as the loss function for the training.
Here, $\mc Q\geq 0$ and $\mc R>0$ are the state and control cost matrices, respectively. Also, $N_s\in \mb N_+$ denotes the training steps. Note that although we consider continuous-time systems in this paper, the simulation is inevitably conducted in discrete time. To be specific, in the present experiment, we choose $\mc Q = I$, $\mc R = 0.1I$ and $N_s = 5000$. We use MuJoCo XLA (MJX) \cite{todorov2012mujoco} to model the plant, where the default sampling period is $h=0.002~\mathrm{sec}$. With this setting, the training horizon is $N_sh = 10~\mathrm{sec}$. In the figures presented below, we reconstruct the continuous time scale.

We set the initial displacements of the masses to be $\begin{bmatrix}
			q_1&q_2&q_3
		\end{bmatrix}=\begin{bmatrix}
			1&2&3
		\end{bmatrix}~\mathrm{meters}$ and calculate the value of the loss function (\ref{eq:loss function J}) by simulating the closed-loop system over the $N_s$ steps of learning horizon. We apply gradient descent to update the parameters of the neural networks in the controller (\ref{eq:controller}) so that (\ref{eq:loss function J}) is minimized.

\begin{figure*}[h!]
\centering
\includegraphics[width=\textwidth]{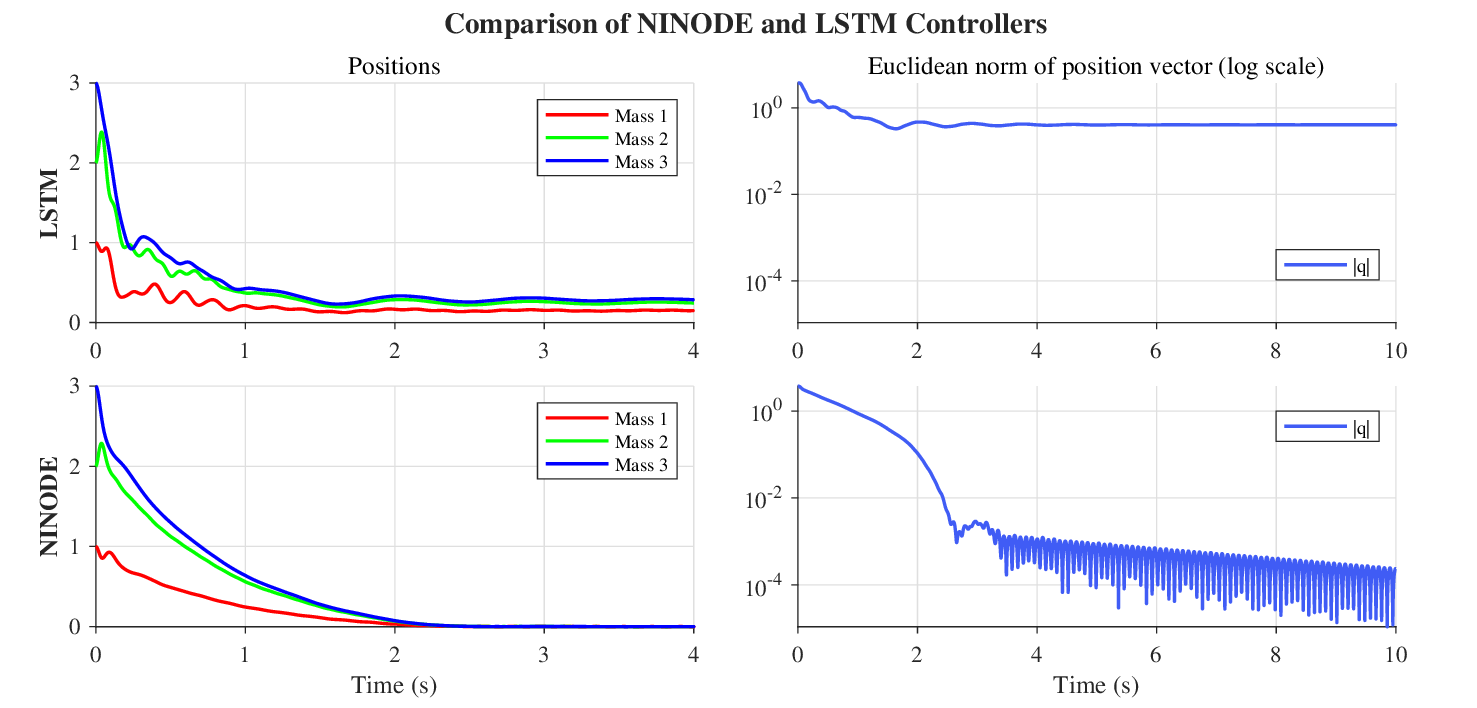}
\caption{Trajectories of the positions under the control of an LSTM controller and the proposed NINODE controller. The left-column figures show the positions of individual masses for $4\ sec$, while the right-column figures show the Euclidean norm of the position vector $q$ in a log scale.}
\label{fig:3 inputs comparison}
\end{figure*}

We also applied a long short-term memory (LSTM) neural network \cite{hochreiter1997long}
controller to the plant. The LSTM controller is also trained to minimize the same cost function (\ref{eq:loss function J}). Fig.~\ref{fig:3 inputs comparison} shows the position trajectories of the masses under the control of these two controllers. The LSTM controller forces the masses to stay close to but not exactly at their equilibrium positions. In contrast, the proposed NINODE controller successfully stabilized the masses to their equilibrium positions. This indicates that the NINODE controller asymptotically stabilized the plant while the LSTM controller failed to do so.

\begin{figure}[h!]
\centering
\includegraphics[width=0.53\textwidth]{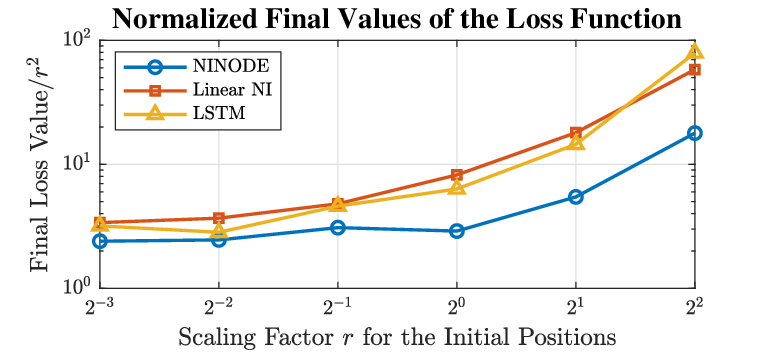}
\caption{Comparison between the NINODE controller, a linear NI controller and an LSTM controller in terms of the normalized final values of the loss function (\ref{eq:loss function J}) after adequate training for different scales of the initial conditions.}
\label{fig:LossComparison_normalized}
\end{figure}

As one might argue that a linear NI controller can also guarantee stability, we illustrate the extra benefits brought by the NINODE controller in the following. We compare the proposed nonlinear controller (\ref{eq:controller}) with its linear version. That is a linear NI controller constructed in a similar way as (\ref{eq:controller}) but with linear matrices $J_2$, $R_2$, $C_2$, and a linear quadratic Hamiltonian $H(x_2)$; i.e., $H_2(x_2) = x_2^TP_2x_2$, where $P_2=P_2^T>0$. We compare the minimum values of the loss function (\ref{eq:loss function J}) that can be achieved after adequate training in the cases in which the NINODE controller, the linear NI controller, and the LSTM controller are used. To be specific, we scale the initial condition with a scaling factor $r=2^{-3}, 2^{-2}, 2^{-1}, 2^0, 2, 2^2$ and find the final values of the loss function (\ref{eq:loss function J}) after training. For each scaling factor $r$, the NINODE controller (\ref{eq:controller}) leads to a smaller final value of the loss function than the other two controllers, as shown in Fig.~\ref{fig:LossComparison_normalized}. Because of the large differences in the achieved final values of the loss function for different scaling factors, we use a log scale for the axis corresponding to the final loss values. It can be observed that the proposed NINODE controller can achieve a much smaller final value of the loss function after training for each of the tested scaling factors.

 \section{Conclusion}
 We propose a novel neural network controller which has intrinsic NI properties. We show that under certain assumptions, such a controller can asymptotically stabilize an NI plant. In particular, with suitable regularity parameters, such a controller can provide guaranteed asymptotic stability to a mechanical plant with colocated force actuators and position sensors. We illustrated the utility of the proposed NINODE controller and its advantages over a linear NI controller and an LSTM controller using a mass-spring example.

\end{document}